\documentclass[prl,aps,twocolumn,psfig,preprintnumbers,amsmath,amssymb,showpacs]{revtex4}
\usepackage{graphicx}\usepackage{epsfig}%\usepackage{portland}
\usepackage{hyperref}
\begin{document}
% \setpagewiselinenumbers
%  \modulolinenumbers[5]
%  
%\preprint{preprintnumber}
\title{
     $\alpha$-clustering effects in dissipative $^{12}$C+$^{12}$C reactions at $95$ MeV
}
%
%   --- AUTHORS
%
\author{G.~Baiocco$^{1,2}$}
\author{L.~Morelli$^{1}$}
\author{F.~Gulminelli$^2$}
\author{M.~D'Agostino$^{1}$}
\author{M.~Bruno$^{1}$}
\author{U.~Abbondanno$^3$}
\author{S.~Barlini$^{4,5}$}
\author{M.~Bini$^{4,5}$}
\author{S.~Carboni$^{4,5}$}
\author{G.~Casini$^5$}
\author{M.~Cinausero$^6$}
\author{M.~Degerlier$^7$}
\author{F.~Gramegna$^6$}
\author{V.~L.~Kravchuk$^{6,8}$}
\author{T.~Marchi$^{6,9}$}
\author{A.~Olmi$^{5}$}
\author{G.~Pasquali$^{4,5}$}
\author{S.~Piantelli$^{5}$}
\author{Ad.~R.~Raduta$^{10}$}

\address{$^1$Dipartimento di Fisica ed Astronomia dell'Universit\`{a} and INFN,
Bologna, Italy}
%\address{$^2$INFN, Bologna, Italy}
\address{$^2$LPC (IN2P3-CNRS/Ensicaen et Universit\'e), 
F-14076 Caen c\'edex, France}
\address{$^3$INFN Trieste, Italy}
\address{$^4$Dipartimento di Fisica ed Astronomia dell'Universit\`{a}, Firenze, Italy} 
\address{$^5$INFN Firenze, Italy} 
\address{$^6$INFN, Laboratori Nazionali di Legnaro, Italy}
\address{$^7$ University of Nevsehir, Science and Art Faculty, Physics Department, Nevsehir, Turkey}
\address{$^8$National Research Center \textquotedblleft Kurchatov Institute\textquotedblright , Moscow, Russia}
\address{$^9$Dipartimento di Fisica ed Astronomia dell'Universit\`{a}, Padova, Italy}
%\address{INFN, Napoli, Italy}
\address{$^{10}$NIPNE, Bucharest-Magurele, POB-MG6, Romania}
%
%   --- ABSTRACT
%
\begin{abstract}
Dissipative  $^{12}$C+$^{12}$C reactions at 
95 MeV are fully detected in charge with the GARFIELD and RCo apparatuses at LNL. A comparison to a dedicated Hauser-Feshbach calculation allows to select events which correspond, to a large extent, to the statistical evaporation of highly excited $^{24}$Mg, as well as to extract information on the isotopic distribution of the evaporation residues in coincidence with their complete evaporation chain. Residual deviations from a statistical behaviour are observed in $\alpha$ yields and attributed to the persistence of cluster correlations well above the $^{24}$Mg threshold for 6 $\alpha$'s decay.
\end{abstract}

 \maketitle
%   \linenumbers

Since the first heuristic proposition of $\alpha$-chains as possible building blocks of even-even nuclei in the late sixties~\cite{ikeda}, the subject of $\alpha$-clustering has been a central issue in  nuclear physics and has even witnessed a gain of interest in recent years~\cite{freer}.
On the theoretical side, highly sophisticated ab-initio calculations have shown pronounced cluster features in the ground state of a large number of light nuclei~\cite{abinitio} as well as in some excited states around the threshold energy of breakup into constituent clusters, showing that cluster correlations are indeed an ubiquitous feature of quantum few-body systems  down to the femtometer scale.  
Concerning experimental research, rotational bands consistent with $\alpha$-cluster structures have been identified in different even-even light nuclei and shown to persist even along their isotopic chains~\cite{freer}. Exotic non-statistical decays of these correlated states have been evidenced in the recent literature~\cite{cluster}.  

 A natural extension of the concept of nuclear clusters concerns nuclear molecules. Molecular states have been seeked for in nuclear reactions since the early days of heavy-ion science. In particular, several interesting resonances 
have been observed in the   $^{12}$C+$^{12}$C reaction in the inelastic~\cite{inelastic} and 
$\alpha$-transfer channels~\cite{transfer}. These studies suggest that  resonant structures persist   in the $^{24}$Mg system up to around 50 MeV excitation energy, a remarkable result as a pure statistical behaviour might be expected due to the extremely high number of available states at such high excitation. Concerning the $\alpha$-transfer channel, experimental results have been reproduced by coupled cluster calculations~\cite{takashina} where the cross section is dominated by a four-cluster ($\alpha+\alpha$)+($\alpha+^{12}$C) state of highly excited $^{24}$Mg around 30 MeV. Because of the remarkable persistence of cluster correlations at high excitation energies, the question naturally arises whether such correlations might affect other more dissipative channels which are typically associated to the formation of a compound nucleus, that is a system whose decay is assumed to be fully decoupled from the reaction entrance channel and governed by purely statistical laws.

 This effect  might be experimentally seen as an excess of cluster production with respect to the prediction of the statistical model, provided that the ingredients of the latter are sufficiently
constrained via experimental data. To this aim, we have performed an  exclusive and (quasi)complete detection of the different decay products 
emitted in $^{12}$C+$^{12}$C dissipative reactions at $95$ MeV, and compared experimental data to a dedicated Hauser-Feshbach code, with transmission coefficients and level densities optimized on the $A\approx 20$ region~\cite{baiocco}. 

We will show that all the observables of dissipative events are fully compatible with a standard statistical behaviour, with the exception of $\alpha$-yields in coincidence with oxygen residues. The observed anomalies are tentatively attributed to clustering effects which appear to survive even in the most dissipative events.

The experiment was performed at the LNL (Laboratori Nazionali di Legnaro), with the $^{12}$C beam provided by the XTU TANDEM accelerator.
The experimental setup was composed by the GARFIELD detector, covering almost completely the polar angular range from 30$^{\circ}$ to 150$^{\circ}$, and the Ring-Counter (RCo) annular detector, centered at $0^{\circ}$ with respect to the beam direction and covering forward laboratory angles in the range $5^{\circ}\leq \theta \leq 18^{\circ}$.
 Details on the apparatus and the experimental techniques can be found in Ref.~\cite{garf,noi}. 
 The combination of the two devices allows a nearly-$4 \pi$  coverage of the solid angle, which, combined with a high granularity, permits to measure the charge, the energy and the emission angles of nearly all charged reaction products, allowing an excellent discrimination of different reaction mechanisms.\\
Events with a single $Z\ge2$ fragment detected in the RCo (identified only in charge) in coincidence with one or more light charged particles (LCP) $Z\le 3$ in GARFIELD are selected. LCP isotopic identification has been performed through Fast and Slow CsI analysis \cite{fs}, with energy thresholds~\cite{footnote} similar to other experimental $4\pi$ devices employing the same identification method \cite{Rivet}.\\
Only events where the total charge of the entrance channel $Z_{det}=12$ is detected are kept for the analysis, unless specified in the text. We expect that most of the selected events should correspond to fusion-evaporation, with the residue detected at forward angles in coincidence with evaporated particles detected at central center-of-mass angles covered by GARFIELD. Experimental data are therefore compared to the predictions of a Monte Carlo Hauser-Feshbach code~\cite{baiocco} for the evaporation of the compound nucleus $^{24}$Mg, at $E^*_{fus}=61.4$ MeV, corresponding to complete fusion source, and filtered through a software replica of the experimental set-up.
The maximum angular momentum for the fused system is assumed $J_{0\ max}=12\ \hbar$, coming from the systematics in PACE4~\cite{PACE}. 
 The comparison of various experimental observables and code calculations
is used to validate the parameterizations of statistical model ingredients implemented in the code. In particular, an exclusive study of completely reconstructed decay chains has been performed, in order to  get an insight into the deviations from a statistical behaviour observed in the decay.

\begin{figure}
\begin{center}
\includegraphics[angle=0, width=1.\columnwidth]{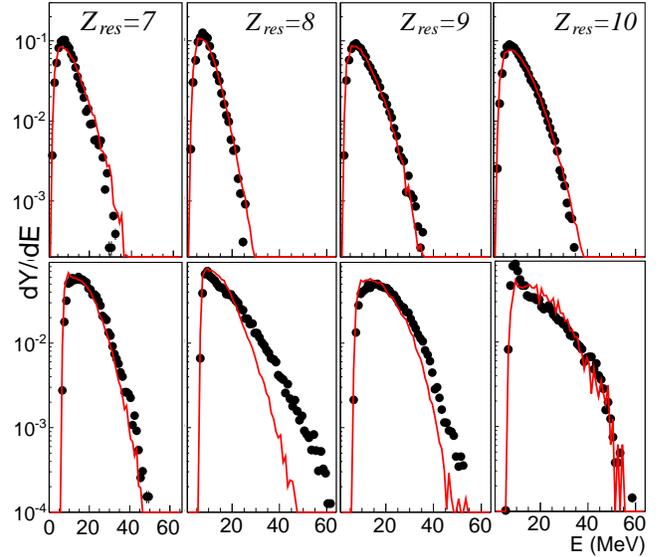}
\end{center}
\caption{(Color online) Proton (upper part) and $\alpha$ (lower part) energy spectra in complete $Z_{det}=12$ events detected in coincidence with a residue of charge $Z_{res}$, indicated in each figure column. Experimental data (dots) are compared to model calculations (red lines). Spectra are shown for the most abuntant residues, $7 \le Z_{res} \le 10$.}
\label{dYdE_p_a}
\end{figure}

Fig.~\ref{dYdE_p_a} displays the energy spectra of protons and $\alpha$ particles detected in GARFIELD in coincidence with the most abundant residues. If not explicitely stated, all distributions are shown normalized to a unitary area, so that a comparison among their shapes is easily done, and all energies are given in the laboratory reference frame. Experimental data are always shown with statistical error bars, when visible.
We can see that a very good reproduction of the proton energy spectra is achieved in all channels, while a large discrepancy in the shape of the distribution appears for $\alpha$ particles
in coincidence with oxygen residues.
The shape of particle spectra produced by a statistical process is determined by the interplay of all physical ingredients entering in the decay, notably transmission coefficients, angular momentum distribution and level density parameters.  Nevertheless, these different ingredients can be largely  disentangled \cite{charity}: transmission coefficients define the shape of evaporated spectra in the Coulomb barrier region; the level density mostly affects the slope of the exponential tail; and the angular momentum steepens the tail of the spectrum of heavy particles ($\alpha$'s) with respect to light ones (protons).
On the basis of such considerations it is found that no common choice on the parameters can be done in our calculations in order to reproduce at the same time protons and $\alpha$ energy spectra. The best model reproduction is obtained with standard fiducial values for the transmission coefficients, level density parameters and angular momentum distribution~\cite{baiocco}, as shown in Fig.~\ref{dYdE_p_a}. The very good reproduction achieved for protons suggests  that the discrepancy found for $\alpha$ particles reflects an out-of-equilibrium emission and mostly concerns well-defined channels where oxygen is produced. 

\begin{figure}
\begin{center}
\includegraphics[angle=0, width=1.\columnwidth]{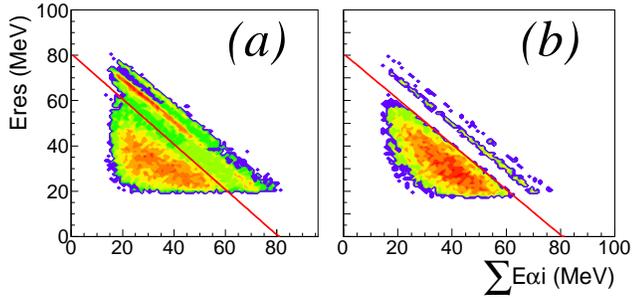}

\end{center}
\caption{(Color online) Experimental (a) and theoretical (b) correlation between the energy of the oxygen fragment and the sum of the energies of the two $\alpha$ particles in $^{12}$C($^{12}$C,$^{A}$O)$\alpha\alpha$ events. The line represent the kinematical locus $Q_{kin}=E_{res}+\sum_{i=1}^{2} E\alpha_i-E_{beam}=-15.78$ MeV at the opening of the 4-body channel $^{12}$C($^{12}$C,$^{15}$O)$n\alpha\alpha$.}
\label{bi_2a_O_exp_cut}
\end{figure}

To understand the deviation, a more exclusive channel by channel comparison is therefore needed. Fig.~\ref{bi_2a_O_exp_cut} displays the experimental (left panel) and theoretical (right) correlation between the energy of the oxygen fragment $E_{res}$ and the sum of the energies of the two $\alpha$-particles $E\alpha_i$  in  $^{12}$C($^{12}$C,$^{A}$O)$\alpha\alpha$ events.
The lines represent the kinematical locus $Q_{kin}=E_{res}+\sum_{i=1}^{2} E\alpha_i-E_{beam}=-15.78$ MeV, where $E_{beam}$ is the beam energy.
This locus divides the ($E_{res}$,$\sum_{i=1}^{2} E\alpha_i$) plane into two regions: in the one above, we find two correlated bands centered respectively at $Q_{kin} = (1.27 \pm 1.40)$ MeV and $(-5.35 \pm 1.70)$ MeV, while a broader region extends below the locus up to a high amount of missing energy. In the following, we will refer to these regions as respectively the non-dissipative and dissipative events region, and we will adopt for them the notation $Q_{>}$ and $Q_{<}$.
In the statistical model interpretation, the two correlated bands correspond to $\alpha$-decay chains, starting from the $^{24}$Mg$^*$ compound nucleus and leaving an $^{16}$O residue either in its ground state ($Q_{kin}= -0.11$ MeV) or in one of its excited bound states which are not resolved in the experiment, with an average energy expense $Q_{kin}\approx-6.55$ MeV. The broadening observed for the experimental $Q_{kin}$ bands reflects the energy resolution of our setup for the energy balance in triple coincidence measurements, which adds up to $\Delta E/E \lesssim 4\%$. 
The locus $Q_{kin}=-15.78$ MeV gives the threshold $Q$-value where the 4-body channel $^{12}$C($^{12}$C,$^{15}$O)$n\alpha\alpha$ opens. In terms of energy, this is the less expensive channel involving an oxygen residue and particles other than $\alpha$'s. Neutrons are not detected in this experiment, and the broader distribution observed for lower $Q_{kin}$ values is due to events in which neutron(s) emission has taken place, and their kinetic energy has not been collected. According to model calculations, the (n,2$\alpha$,$^{15}$O) channel should absorb the largest cross-section in this region.

Besides the common pattern observed for the (2$\alpha$,$^{A}$O) energy correlation, a difference in the relative population of the $Q_{\lessgtr}$ regions is evident between experimental data and calculations in Fig.~\ref{bi_2a_O_exp_cut}. 
In particular, a much higher percentage of (2$\alpha$,$^{A}$O) events populates the $Q_{>}$ region in the experimental sample: (37$\pm$5)\%, with respect to (9$\pm$1)\% according to model predictions. Also in the following, when giving experimental percentages, the associated error is an upper limit taking into account both the
statistical error and the possible $^3$He-$\alpha$ contamination. %CHECKK
Events falling in the $Q_{>}$ region correspond to exit channels with low energy dissipation. Therefore, this region can also be populated by reactions not proceeding through an intermediate compound nucleus state. The larger experimental branching ratio for the (2$\alpha$,$^{16}$O) exit channel in the $Q_{>}$ region probably reflects a  contamination of direct ($\alpha$-transfer/pick-up) reactions, in competion with fusion-evaporation. \\

\begin{figure}
\begin{center}
\includegraphics[angle=0, width=1.\columnwidth]{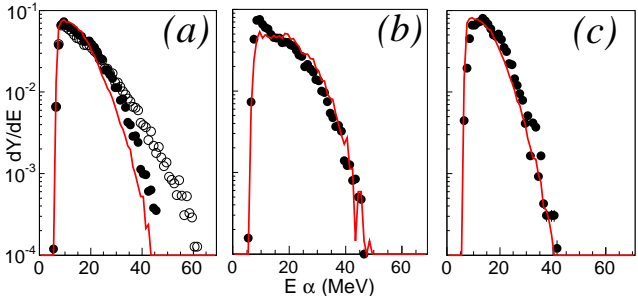}
\end{center}
\caption{(Color online) For dissipative events (as defined in the text) with $Z_{det}=12$, experimental (black dots) and calculated (red lines) $\alpha$ energy spectra in coincidence with an oxygen are compared for: all channels involving at least 1$\alpha$ (a) and, separately, channels involving $2\alpha$'s (b) and only 1 $\alpha$ (c). As a reference, in the leftmost panel, also the $\alpha$ energy distribution without any $Q_{kin}$ selection (same as in Fig.~\ref{dYdE_p_a}) is shown (empty dots).}
\label{dYdE_a_Z8_cut_new}
\end{figure}

Fig.~\ref{dYdE_a_Z8_cut_new}\textit{a} displays the comparison between data and calculations for the energy spectrum  of $\alpha$ particles detected in coincidence with an oxygen residue and emitted in dissipative events (full dots). As a reference the $\alpha$ energy distribution without any $Q_{kin}$ selection (empty dots, see Fig.~\ref{dYdE_p_a}) is also plotted.
A great improvement in the agreement between data and calculation is achieved when we limit the comparison to the $Q_{<}$ event class. Nevertheless, a residual discrepancy is observed, and the experimental distribution is still not reproduced in its shape.
To investigate the origin of this deviation, in the middle (\textit{b}) and right (\textit{c}) panels of Fig. \ref{dYdE_a_Z8_cut_new} we show the $\alpha$ energy distributions associated to channels where the oxygen residue is in coincidence with, respectively, two or only one $\alpha$('s).
We can see that the shape of the energy spectra depends on the channel. The statistical calculation is able to satisfactorily reproduce separately the energy distributions of $\alpha$ particles for dissipative $2\alpha$'s and $1\alpha$ completely reconstructed decay chains, while the sum of the two processes is not correctly reproduced. Experimentally,  (63$\pm$5)\% of the total cross section  for completely reconstructed dissipative decays with a $\alpha$ particle and an oxygen fragment in the outgoing channel is absorbed by (2$\alpha$,$^{16}$O$^*$) channels, where the oxygen is excited above its neutron emission threshold, while according to the theoretical predictions these channels should represent only (10$\pm$1)\% of this class of events. In conclusion, the residual deviation observed in the leftmost panel of Fig.~\ref{dYdE_a_Z8_cut_new} appears to originate from a branching ratio discrepancy for the ($n$,2$\alpha$,$^{15}$O) (and, with a minor contribution, for the  $2n$,2$\alpha$,$^{14}$O)) decay channel
with respect to decay channels involving a single $\alpha$ emission, even if a contribution from high energy $\gamma$ emission from collective states cannot be a priori excluded.

\begin{figure}
\begin{center}
\includegraphics[angle=0, width=\columnwidth]{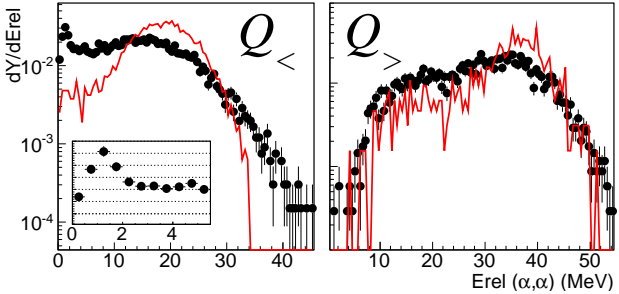}
\end{center}
\caption{(Color online) Experimental (black dots) and calculated (red lines) relative energy distributions of the two $\alpha$'s found in coincidence with an oxygen in dissipative (left) and non-dissipative (right) events. In the left panel, a zoom on the low relative energy region is shown in the figure inset for experimental data.}
\label{correlation}
\end{figure}

This finding indicates that cluster correlations associated to the $^{12}$C+$^{12}$C system persist up to higher center-of-mass energies than previously expected from the analysis of the inelastic channel~\cite{bremner}, and lead to a non-statistical behaviour in the decay of the highly excited $^{24}$Mg. 
Given the high excitation energy $E^*=61.4$ MeV it is not possible to associate the extra yield to a single isolated state with a well defined angular momentum and parity. Nevertheless, interesting
information can be extracted from Fig.~\ref{correlation}, which displays
the $\alpha-\alpha$ relative energy distribution for the $Q_<$ and the $Q_>$ region, in comparison with the statistical model calculation. We can see that a clear peak appears in the most dissipative events, which can be associated to doorway ($^8$Be-$^{16}$O$^{*}$) and ($^9$Be$^*$-$^{15}$O) states,
where the excitation energy of the $^{16}$O ($^9$Be) is above the
neutron emission threshold, and which is not present in
the statistical model. The absence of a $^{8}$Be resonance in the less dissipative events suggests that these reactions do not originate from a ($\alpha+\alpha$)+($\alpha+^{12}$C) doorway state as it was suggested to be the case at lower energies \cite{takashina}.

\begin{figure}
\begin{center}
\includegraphics[angle=0, width=\columnwidth]{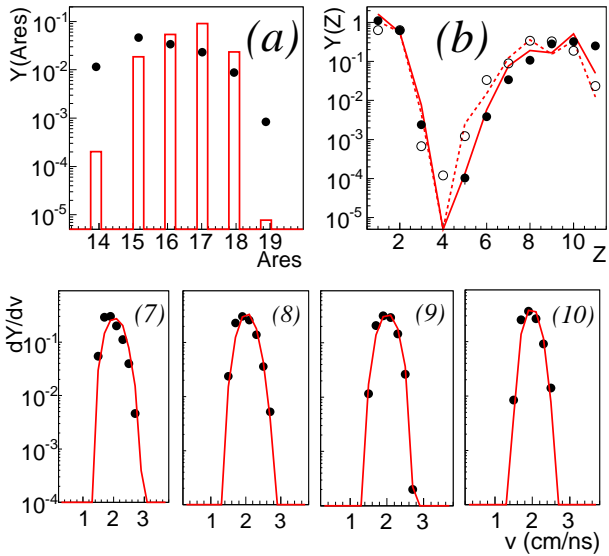}
\end{center}
\caption{(Color online) Panel (a): the experimental oxygen isotopic distribution (reconstructed via the minimization of the quantity $\delta\varepsilon^*_{cal}$ defined in the text, black dots) is compared to the model predictions (red histograms) for all events. 
Panel (b) and lower part, respectively: charge distribution of reaction products and velocity distributions of most abundant evaporation residues (charge $Z_{res}$ indicated in parenthesis) are compared for data (dots) and model calculations (lines) after the  
exclusion of non-dissipative events.
For the charge distribution, experimental data for complete ($Z_{det}=12$, black points) and quasi-complete ($Z_{det}\ge10$, empty points) events
are compared to model calculations (solid/dashed lines respectively).
Mass and charge distributions are normalized to the number of events retained in the analysis.}
\label{all}
\end{figure}

Since we have interpreted the observed extra $\alpha$-yield from dissipative events as associated 
to the production of neutron-poor oxygen isotopes, we expect that a signature of the anomalous 
branching ratio should be evidenced in  the isotopic residue distribution. Unfortunately the residue mass is not directly accessible in this experiment, because of the low energy of heavier fragments. However, due to the completeness of the detection, this information can be approximately deduced from the calorimetric energy distribution, as explained hereafter.
The excitation energy  $E^*_{fus}$ can be 
estimated event-by-event from the calorimetric energy balance $E^{*}_{cal}(A_{res})$,
provided an hypothesis is made on the residue mass number $A_{res}$:
\begin{equation}
E^{*}_{cal}(A_{res})=
\sum_{i}^{Nc}E^{CM}_{ i}+N_{n}(A_{res})\cdot\left\langle E^{CM}_{n}\right\rangle 
+Q(A_{res})
\label{calorimetryeq}
\end{equation}
where $N_c$ ( $N_{n}$ ) and $E^{CM}_{ i}$ ($\left\langle E^{CM}_{n}\right\rangle$) are respectively the charged products (neutron) multiplicity and their center-of-mass kinetic energies, and $Q$ is the decay $Q$-value. 
Indeed, the emitted neutron number is uniquely determined by mass conservation and their energy can be (in average) estimated from the measured proton energy with the subtraction of the Coulomb barrier.
An estimation of the  mass of the residue can thus be obtained 
by minimizing in each event completely detected in charge the quantity $\delta\varepsilon^*_{cal}(A_{res})= \vert E^{*}_{cal}(A_{res})-E^*_{fus}\vert$.
Fig.~\ref{all}\textit{a} displays the resulting  estimated oxygen isotopic distribution. 
As expected, we find in the data an extra production of neutron-poor isotopes, coherent with the finding of an extra experimental cross section for the ($xn$,$2\alpha$,$^{16-x}$O) channel. The reconstructed experimental isotopic distribution for the considered residue is found to be generically wider with respect to model predictions.  

Finally, it is important to stress that, with the exception of the anomalous branching ratio towards this channel, the detected events where the dissipated energy overcomes the four-body threshold associated to $^{15}$O formation ($Q_{kin}=-15.78$ MeV) are fully compatible with a complete fusion pattern followed by the compound nucleus decay of $^{24}$Mg$^*$. Indeed the Hauser-Feshbach calculation satisfactorily reproduces all the  different observables that can be constructed  with the exclusion of $Q_{>}$ events from the analysis, and that do not directly involve $\alpha$-oxygen coincidences.
Fig.~\ref{all} displays some selected examples, as the velocity distributions of the heaviest fragment and the charge distribution of reaction products, this latter both for complete ($Z_{det}=12$) and almost-complete ($Z_{det}\ge10$) events, in order to show that no important bias is induced by the stringent completeness conditions. The global quality of the agreement between data and calculations makes us claim that we have been able to select in our data-set a component which is as close as possible to the statistical decay of a hot equilibrated source, corresponding to the complete fusion $^{24}$Mg$^*$ source.

In conclusion, in this paper we have presented  an exclusive analysis of dissipative $^{12}$C+$^{12}$C collisions at $95$ MeV, fully reconstructed in charge. A detailed comparison to a dedicated Hauser-Feshbach calculation shows an abnormally high branching ratio towards  the (2$\alpha$,$^{16}$O$^*$) channel with respect to the statistical expectation, where the oxygen is excited above its neutron emission threshold, which corresponds in part to the population of a doorway ($^8$Be-$^{16}$O$^*$) configuration.
This extra yield could be due to a failure of the Hauser-Feshbach theory for the decay of highly excited $^{24}$Mg, possibly due to the $\alpha$-correlations of this nucleus which are already suspected to lead to specific high-lying ($^{16}$O-$2\alpha$) resonances~\cite{kawabata}. Alternatively, the compound nucleus hypothesis of full decoupling between entrance and exit channel could be questioned due to $\alpha$ correlations in the   $^{12}$C+$^{12}$C entrance channel. These two hypotheses cannot be discriminated by means of the present experimental information, but new data on $^{14}$N+$^{10}$B at 80 MeV are currently under analysis to give a definitive answer to this issue.\\
\ \\
\textit{Acknowledgements}\\
The authors would like to thank the crew of the XTU TANDEM acceleration system at LNL for the smooth running of the machine, and, in particular, for the fruitful collaboration with the scientific and technical coordinators.\\
This work was partially supported by the European Funds for Large Scale Facilities - Seventh Framework Program - ENSAR 262010.


\begin{thebibliography}{99}
\bibitem{footnote} The energy thresholds were 2, 6, 14, 20, 8 MeV respectively for $p$, $d$, $t$, $^3$He and $\alpha$ particles. The $^3$He can be discriminated from $\alpha$'s starting from 20 MeV. This increases the $^3$He threshold but does not affect too much the $\alpha$ yield since $^3$He is estimated to represent only less than 2-3\% of Z=2 particles.
\end{thebibliography}
\end{document}